\documentclass[12pt]{article}
\usepackage{amsmath,epsf,amssymb,latexsym,cite,eucal}

\newif\iffigs\figstrue

\DeclareFontFamily{U}{rsf}{}
\DeclareFontShape{U}{rsf}{m}{n}{
  <5> <6> rsfs5 <7> <8> <9> rsfs7 <10-> rsfs10}{}
\DeclareMathAlphabet\Scr{U}{rsf}{m}{n}

\def\Q{{\mathbb Q}}
\def\R{{\mathbb R}}
\def\Z{{\mathbb Z}}

\def\Im{\operatorname{Im}}

\def\Area{\operatorname{Area}}
\def\Vol{\operatorname{Vol}}

\def\GU{\operatorname{U{}}}

\def\ch{\operatorname{\mathit{ch}}}
\def\td{\operatorname{\mathit{td}}}

\def\Re{\operatorname{Re}}

\def\CY{Calabi--Yau}

\def\ff#1#2{{\textstyle\frac{#1}{#2}}}
\def\half{\frac{1}{2}}

\def\N{{\mathcal{N}}}

\def\eqn#1#2{\begin{equation}#2\end{equation}}

\begin{document}

\begin{titlepage}
\begin{flushright}
DUKE-CGTP-06-01\\
SLAC-PUB-12048\\
October 2006
\end{flushright}
\vspace{.5cm}
\begin{center}
\baselineskip=16pt
{\fontfamily{ptm}\selectfont\bfseries\huge
Black Hole Entropy, Marginal Stability\\[3mm] and Mirror Symmetry}\\[20mm]
{\bf\large  Paul S.~Aspinwall$^{1}$, Alexander Maloney$^{2}$ and Aaron
Simons$^{3}$
 } \\[7mm]

{\small

$^1$ Center for Geometry and Theoretical Physics,
  Box 90318 \\ Duke University,
 Durham, NC 27708-0318 \\ \vspace{6pt}
$^2$ School of Natural Sciences, Institute for Advanced Study \\
Einstein Dr., Princeton, NJ 08540 \\ \vspace{6pt} $^3$ Department of
Physics, Harvard University, Cambridge, MA 02138 \\ \vspace{6pt}

 }

\end{center}

\begin{center}
{\bf Abstract}
\end{center}
We consider the superconformal quantum mechanics associated to BPS
black holes in type IIB Calabi--Yau compactifications. This quantum
mechanics describes the dynamics of D-branes in the near-horizon
attractor geometry of the black hole.  In many cases, the black hole
entropy can be found by counting the number of chiral primaries in
this quantum mechanics. Both the attractor mechanism and notions of
marginal stability play important roles in generating the large
number of microstates required to explain this entropy. We compute
the microscopic entropy explicitly in a few different cases, where
the theory reduces to quantum mechanics on the moduli space of
special Lagrangians.  Under certain assumptions, the problem may be
solved by implementing mirror symmetry as three T-dualities: this is
essentially the mirror of a calculation by Gaiotto, Strominger and
Yin. In some simple cases, the calculation may be done in greater
generality without resorting to conjectures about mirror symmetry.
For example, the $K3 \times T^2$ case may be studied precisely using
the Fourier-Mukai transform.

\vspace{2mm}
\vfill
\hrule width 3.cm
\vspace{1mm} {\footnotesize \noindent email: psa@cgtp.duke.edu, \,
  maloney@ias.edu, \, simons@physics.harvard.edu}
\end{titlepage}

\vfil\break


\section{Introduction}    \label{s:intro}

\def\M{{\mathcal{M}}}
\def\A{{\mathcal{A}}}
\def\F{{\mathcal{F}}}
\def\J{{\mathcal{J}}}

The study of quantum mechanical black holes continues to provide new
insights into the basic structure of string theory and quantum gravity.
In this paper we will focus on supersymmetric black holes arising in
Calabi--Yau compactifications of type II string theory, which
are composed of BPS D-branes wrapping various cycles in the Calabi--Yau.

The most straightforward computations of black hole entropy involve
a direct enumeration of supersymmetric ground states of particular
D-brane system \cite{Strominger:1996sh}.
In general this computation is quite difficult,
but considerable progress has
been made in a variety of special cases, often by using
duality, as in \cite{Maldacena:1997de}.
Recently, however, an alternative approach to black hole entropy has
been proposed \cite{GSSY:D0,GSY:bqm}.
Rather than study the complete
brane configuration, one instead investigates
the near horizon quantum mechanics of a collection of D-branes
moving in the supergravity background geometry sourced by the remainder
of the branes comprising the black hole.
By separating the system into
``probe'' and ``background'' branes in this way, many of the mathematical
problems become considerably more tractable.
One can then, at least in some cases, reproduce entropy by enumerating
the supersymmetric ground states of the near horizon probe quantum mechanics.

Actually, as we will argue below, this picture of ``probe'' and
``background'' is essentially unavoidable once one realizes that the
attractor mechanism forces BPS black holes to be marginally stable.

In this paper we will adopt the strategy of \cite{GSSY:D0,GSY:bqm}.
These papers focused on type IIA, which contains a variety of
BPS states: the B-branes, which --- very roughly speaking ---
correspond to holomorphic sub-manifolds of the \CY.
The authors of \cite{GSSY:D0,GSY:bqm} successfully
reproduced the entropy of black holes whose charge is comprised mainly of
0-branes.  In doing so, they used the fact that
the moduli space of a probe 0-brane on the \CY\ is just the \CY\ itself.
So the near horizon theory of a collection of probe 0-branes reduces
to a non-linear sigma model whose target space is the Calabi--Yau.
The general case --- which involves
the quantum mechanics of various higher dimensional
branes --- remains unsolved,
although some progress has been made \cite{Gaiotto:2005rp}.
In this paper we will focus on type IIB,
where we are presented with only one type of BPS state:
the A-branes, which wrap
special Lagrangian 3-cycles of the Calabi--Yau.
One might therefore hope that a IIB description would allow us to compute
in one fell swoop the black hole entropy for all possible BPS configurations.
In this paper we will report only partial progress towards this
ambitious goal.

To start, we must first understand the quantum mechanics on the moduli
space of A-branes. The analysis of the moduli space is somewhat
subtle.  This is because as the \CY\ moduli are varied, it is possible
for these D-branes to decay and form bound states with other D-branes.
In fact, the attractor mechanism forces the moduli to take on special
values which happen to make the D-brane marginally stable against a
large number of decays.  This leads one to consider the moduli space
of decay products. In general, one can then enumerate the quantum
states of a D-brane black hole by the following procedure
\begin{enumerate}
\item Enumerate the possible marginal decay products.
\item Enumerate the marginal bindings ``at threshold'' of these products.
\item Calculate the cohomology of the moduli space of the resulting
  objects.
\end{enumerate}
In terms of string coupling, this first step is classical while the
second and third steps are quantum mechanical. For the first step
one may use concepts such as $\Pi$-stability or the existence of
special Lagrangians. The second step involves the use of the Myers
effect \cite{Myers:1999ps}, following \cite{GSSY:D0, GSY:bqm}.

Most of this paper will focus on the third step, where we compute
cohomology on the moduli space of special Lagrangian 3-brane probes.
The crucial point is that the RR charge of a black hole background
interacts with the D-brane probe, effectively producing a ``magnetic
field'' on the moduli space of the D-branes probe. That is, the
cohomology computation is bundle-valued for some $\GU(1)$-bundle. We
will argue that $c_1$ of this bundle is proportional to the K\"ahler
form on moduli space, and that moreover this K\"ahler form is fixed
by the attractor mechanism. This is very similar in flavor to the
computation of \cite{GSY:bqm}. These authors focused on D0-D4
systems but we will attempt, with modest success, to be more
general.

The primary problem is then to determine the structure of the
A-brane moduli space.
In general this is quite complicated, but in
some cases one can use mirror symmetry to turn this back into
a problem in type IIA.
This relies on the conjecture of \cite{SYZ:mir} that any Calabi--Yau
manifold can be written as a (possibly degenerate) $T^3$ fibration.
If we consider black holes whose charge comes mostly from
D3 branes wrapping this $T^3$ fiber, then we may use the techniques of
\cite{SYZ:mir} to study this moduli space.
This argument reproduces the correct entropy, and
is essentially the mirror of the computation of \cite{GSY:bqm}.

This argument suffers from the same problems as the
original SYZ construction.  In particular, any proper argument must account
for the degeneracies of the $T^3$ fibration.  The remainder of the paper
describes the more precise $K3 \times T^2$ construction,
which uses the Fourier-Mukai transform to implement mirror symmetry.

An outline of this paper is as follows. In section \ref{s:decay} we
discuss attractive \CY\ three-folds, and demonstrate that they
have special properties when it
comes to D-brane decay. In section \ref{s:slag} we
analyze the quantum mechanics of D-branes wrapping special Lagrangian
cycles.
In section \ref{s:0B} we perform an entropy
calculation by implementing mirror symmetry as T-duality.
As an illustration of this technique, we describe the special case
of type IIB on $T^6$.
In section \ref{s:K3} we describe an analogous computation for
compactifications on $K3 \times T^2$. In this case
we can do the computation exactly, without resorting to conjectures
about the action of mirror symmetry.  We end with a few concluding remarks.


\section{Black Hole Attractors and D-Brane Decay}  \label{s:decay}

We start by reviewing the black hole attractor mechanism
in type IIB, before describing its relation to marginal stability and
D-brane decay.

\subsection{Review: IIB Attractors} \label{ss:attr}

Consider type IIB string theory compactified on a Calabi--Yau 3-fold $Y$.
In the perturbative string description, a supersymmetric state of
this theory is given by a
D3 brane wrapping a special Lagrangian 3-cycle of $Y$.
Such a wrapped D3 brane looks like
a charged point-like object in four dimensions, whose charge depends on the
choice of 3-cycle.
We will denote by $F^{(3)}$ the element of $H^3(Y,{\Z})$ characterizing
this charge: $F^{(3)}$ is
Poincar\'e dual to the homology cycle of the D3 brane.

We can also describe this supersymmetric object as a charged BPS black
hole solution of $\N=2$ supergravity in four dimensions
\cite{FKS:bh,Strominger:1996kf,Ferrara:1996dd}.
This solution exhibits a curious feature known as
the attractor mechanism: at the horizon of the black hole
the vector multiplet moduli approach fixed values, which are determined
only by the charge $F^{(3)}$ and not by the asymptotic values
of the moduli.
In type IIB, these moduli describe complex structure deformations of $Y$.
At the horizon, the moduli are fixed by the condition that
$F^{(3)}$ lies in $H^{3,0}(Y) \oplus H^{0,3}(Y)$
\cite{Moore:arith}.
The holomorphic three form $\Omega$ on $Y$ is a basis element of
$H^{3,0}(Y)$, so this condition can be written as
\begin{equation}
F^{(3)} = \textrm{Im}(C\Omega) \label{eq:att}
\end{equation}
for some complex constant $C$. By introducing a
symplectic basis for $H^3(Y,{\Z})$, one can write this as an
equation for the periods of the holomorphic 3-form.  However,
(\ref{eq:att}) is sufficient for our purposes.

This attractor equation (\ref{eq:att}) is an equation for
$2 h_{2,1}+2$ unknowns --- the $2h_{2,1}$ complex structure moduli and
the complex constant $C$ --- in terms of the
$b_3(Y)=2h_{2,1}+2$ charges.
So it is natural to expect that solutions of (\ref{eq:att}) are isolated
as points in moduli space.  However, it has been shown that
although the solutions are isolated
they are not always unique \cite{Moore:arith}.

To describe the structure of these solutions further, note that
$\N =2$ supergravity contains a gauge boson in the
gravity multiplet --- the graviphoton --- in addition
to gauge bosons in vector multiplets.  The charge measured by the graviphoton
plays a special role in the solution, since it is the
central charge appearing in the supersymmetry algebra.
It is this charge that appears in all BPS-type relations describing
the black hole solutions.  In terms of $F^{(3)}$, this charge is
\eqn\cenc{Z=i e^{K/2} \int \Omega \wedge F^{(3)},}
where
\eqn\kpt{e^{-K}=i\int \Omega \wedge \bar{\Omega}}
is the K\"ahler potential on the vector multiplet moduli space.
The constant appearing in (\ref{eq:att}) can be fixed by wedging
both sides with $\Omega$ and integrating over $Y$.  It is
$C = 2 {\bar Z} e^{K/2}$.

The near horizon geometry of the black hole is
$AdS_2 \times S^2 \times Y$, where the geometry of $Y$ is constrained by
(\ref{eq:att}).  The $AdS_2$ and
$S^2$ factors both have radius $|Z|$ in four dimensional Planck units.
So the Bekenstein-Hawking entropy,
which is proportional to the area of the $S^2$, is
\eqn\asd{
S = \pi |Z|^2 .
}
In this formula the central charge
$Z$ is evaluated at the attractor fixed point
(\ref{eq:att}).

In addition, the D3 brane sources a 5-form field strength, which at
the horizon is \eqn\fsour{F^{(5)}=\omega_{AdS_2} \wedge F^{(3)} +
\omega_{S^2} \wedge \star_6 F^{(3)}. \label{eq:fsour}} Here
$\omega_{AdS_2}$ and $\omega_{S^2}$ are volume forms on $AdS_2$ and
$S^2$, and $\star_6$ is the Hodge star on $Y$.

The attractor solutions we have just described are valid only
when certain conditions are met.
First, in order for the supergravity approximation to be good
the characteristic length scale of $AdS_2\times S^2$ must
be large. That is, the area of the event horizon must be large
compared to the string scale.  So we must take $|Z| \gg 1$, which
can be accomplished, for example, by taking all of the charges to be large.

The second condition is slightly more subtle. We are assuming the
degrees of freedom observed are that of a compactification on $Y$.
That is, we are ignoring any massive excitations of the \CY\ threefold
(although in principle these may be accounted for in the context of the
attractor mechanism, see e.g. \cite{Hsu:2006vw}). This
requires all the characteristic sizes of $Y$ to be small compared
to the size of $AdS_2\times S^2$. Normally one thinks of ``size'' as being
associated with the complexified K\"ahler form $B+iJ$ of a threefold,
while the deformations of complex structure are associated purely to
the ``shape''. This is a little na\"\i ve, however, as we now discuss.

Let $X$ be mirror to $Y$ and consider deformations of the complex
structure of $Y$ and the mirror deformations of $B+iJ$ of $X$. There
is a partition function of string states associated to these spaces
which will vary with the moduli and respect mirror symmetry. If a
characteristic length in the \CY\ gets large one would expect the
partition function to contain light states. The areas of holomorphic
curves in $X$ are determined by $B+iJ$ and are insensitive to
deformations of complex structure. Thus, if $B+iJ$ has any large
component one would expect the appearance of light states
irrespective of the complex structure.

Mirror to this statement, one expects that there must be complex
structures for which $Y$ exhibits light states irrespective of the
K\"ahler form. This seems counterintuitive at first, as one can rescale
the metric on $Y$ (which is a deformation of the K\"ahler form) to
make all lengths small and thus remove any light (non-massless) states.
However, this argument is too classical. If $Y$ is at ``large complex
structure'' the characteristic length scales within $Y$ will differ
wildly. The canonical example is that of a $T^2$ with one very long and
one very short 1-cycle. If we try to shrink the metric to shorten the
longer scales, we will ``run out of moduli space'' before the offending
light modes can be
brought under control. That is, the shorter lengths would be made so
short that they would violate any ``minimum distance'' constraint as
in \cite{AGM:sd}.

So, our second constraint is that $Y$ should not have large $B+iJ$
(with respect to the horizon area) and that it should not be mirror
to space with large $B+iJ$.  The first of these conditions is not
relevant to black hole solutions, as the $B+iJ$ moduli are contained
in hypermultiplets which are constants for these solutions. So we
are free to choose their values to be whatever we like. The second
condition is a constraint on complex structure moduli, and limits
the charges that we may consider.  In particular, as we will see
later in sections 4 and 5, it will force us to take certain {\it
ratios} of charges to be large.  This constraint is easiest to
understand in the mirror IIA language, where it is just the
requirement that volumes of two cycles on X must be small compared
to the characteristic length scale of the four dimensional black
hole geometry.

\subsection{Marginal Stability at Attractor Points}

A central concept when discussing entropy is the notion of the moduli
space of a D-brane. When constructing a moduli space, the notion of
{\em stability\/} is very important.

In general, many moduli space constructions run as follows: one looks for
the moduli space of some object by constructing the moduli space of more
easily defined objects which satisfy an appropriate stability criterion.
For example, the moduli space of vector bundles with an
Hermitian--Yang--Mills connection is studied by starting with holomorphic
vector bundles and imposing $\mu$-stability.  In discussions of this form,
a special case must always be made for the marginally stable object. A
marginally stable object must be viewed as a ``direct-sum'' of its
constituents in order to obtain the correct moduli space. There might be
other configurations of the constituents which are not equivalent to a
direct sum, but they are considered to be ``S-equivalent'' to the direct
sum and are not counted as different states. We refer to
\cite{Sharpe:subK,HL:mods}, for example, for a discussion of
S-equivalence. The important point is that we need to know if a D-brane is
unstable, marginally stable, or truly stable in order to compute the
moduli space correctly.

The attractor behavior described above has several striking consequences,
which we will exploit in our computation of black hole entropy.
The most crucial of these involves D-brane decay and marginal stability,
as we will now describe.  We will continue to work with type IIB on a
Calabi--Yau $Y$.  For other discussions of special Lagrangians
at attractor points, see \cite{Dnf:Dstab,Denef:dis}.

To understand D-brane decay,
consider what happens to a supersymmetric D3 brane as one varies
the complex structure of $Y$.  This brane wraps
a special Lagrangian submanifold, which is defined as a Lagrangian
submanifold $L\subset Y$ with
\begin{equation}
  dV_L = R\exp(-i\pi\xi)\Omega|_L.
\end{equation}
Here $dV_L$ is the volume form on $L$, $\Omega$ is the holomorphic
3-form on $Y$ and $R$ and $\xi$ are real numbers.
The phase $\xi$ of the special Lagrangian is
constant over $L$. Since $\Omega$ is defined only up to multiplication by an
overall constant, one can set
$\xi=0$ for a given brane.  However, we will need to compare values of
$\xi$ for different branes, so will leave $\xi$ unfixed.
Two D3 branes, wrapping different special Lagrangians,
are mutually BPS only if their respective values of $\xi$ are equal.

Now, as one varies the complex structure of $Y$ it is possible for
the special Lagrangian $L$ to become ``pinched''; that is, at a particular
value of the complex structure moduli $L$ will become the union
of two submanifolds touching at a point \cite{Jc:VsLag}.
At this point the D-brane is marginally stable.
As one deforms the complex structure
past this point, $L$ splits up into two distinct
components $L\to L_1 +L_2$.  In general the phases of the two
components will become distinct, and so the
union $L_1 \cup L_2$ is no longer itself special Lagrangian.  The resulting
pair of D-branes is no longer mutually BPS.
In this way, a single A-brane can ``decay'' $L\to L_1+L_2$ as one
deforms the complex structure.

The key idea in looking for D-brane decays is to find
sub-branes\footnote{The idea of a sub-brane is actually poorly-defined
  but we will use this language here. More correctly one should use
  the language of triangulated categories as explained in \cite{me:TASI-D}.}
$L_1$ into which $L$ can decay.
The decay can occur, i.e. $L$ will be marginally unstable,
at the point in moduli space where $\xi=\xi_1$. That is, when the phases
of two periods of the holomorphic 3-form become equal:
\begin{equation}
  \arg\int_L \Omega = \arg\int_{L_1} \Omega.
\end{equation}
At a generic point in moduli space we expect the periods of $\Omega$ to be
transcendental complex numbers. So
any two A-branes whose charges are not
proportional will typically have different phases. Thus at a
generic point there are no marginally stable D-branes: all
branes are either properly stable or properly unstable.

The attractor fixed points described above are very special, however,
in that they admit many marginally stable branes.
In fact, they admit the
{\it maximal\/} number of marginally stable branes.

To see this, consider the 3-brane black hole described above.  The
D-brane under consideration, with charge $F^{(3)}$,
wraps a special Lagrangian with phase
\eqn\asd{
\arg\int_Y \Omega \wedge F^{(3)}= \arg Z
.}
Consider a second ``probe'' 3-brane with charge $v \in H^3(Y, {\bf Z}) $.
The phase of this A-brane will be aligned (or anti-aligned)
with the phase of the black hole only if
\begin{equation}
\arg(\pm Z) = \arg \int_Y \Omega \wedge v .
\label{eq:marcond}
\end{equation}
Since $v$ is real, this can be written as
\begin{equation}
\begin{split}
0 &= {\bar Z} \int_Y \Omega \wedge v - Z \int_Y {\bar \Omega} \wedge v
= -i e^{-K/2} \int_Y F^{(3)} \wedge v .
\end{split}
\label{eq:align}
\end{equation}
The final expression
is the natural symplectic inner product on $H^3(Y, {\bf Z})$.
Our condition is just that $v$ is perpendicular to $F^{(3)}$ with respect to
this inner product:
$\langle F^{(3)}, v\rangle = \int_Y F^{(3)} \wedge v = 0$.

This is a very striking result. It shows that a D-brane in an
attractor background is far more likely to be marginally stable
than a generic D-brane. In particular, the whole codimension-one sublattice of
$H^3(X,\Z)$  orthogonal to $F^{(3)} $ gives
sub-branes with respect to which $L$ can be marginally unstable.
This is in stark contrast to the generic values of complex structure
discussed above.

It is easy to show that all possible D-brane charges cannot correspond
to mutually BPS states, and so, in this sense, this codimension one
sublattice is the maximal set of states that can be mutually BPS. In
other words, {\em the attractor equations force the D-brane to be
  maximally marginally-stable.}


Thus the attractor mechanism forces us to consider the D-brane as
a ``direct sum'' of constituent objects when we consider moduli spaces.
So at the level of moduli spaces we regard the constituent D-branes as
completely non-interacting. As we will see, however, the RR-fields
{\em do\/} produce interactions between the constituent D-branes and
it is very important to take this into account in order to correctly
compute the entropy of the system.

Although we have focused on A-branes of type IIB, one can make analogous
comments concerning the B-branes in type IIA string theory on the mirror
Calabi--Yau $X$. As B-brane computations tend to be more
tractable, any entropy calculation
will almost certainly be easier in the IIA language.
However, as the mathematical machinery is more abstract we will
make only a few comments here.

A B-brane $E$ may be regarded, in ascending order of honesty,
as a vector bundle over a
holomorphic submanifold, as a coherent sheaf, or as an object in the
derived category of coherent sheaves \cite{Doug:DC}
(see \cite{me:TASI-D} for a review). Its charge is
\begin{equation}
  \ch(E)\wedge\sqrt{\td(T_X)} \in H^{\textrm{even}}(X,\Q).
\end{equation}
The natural inner product between B-brane charges, which is mirror to the
intersection form on 3-cycles given above, is
\begin{equation}
  \langle E, F\rangle = \int_X\ch(E)^\vee\wedge\ch(F)\wedge\td(T_X),
  \label{eq:Binr}
\end{equation}
where ${}^\vee$ reverses the sign of $(4n+2)$-forms for all $n$.
The notion of a stable special Lagrangians is replaced by $\Pi$-stability and
distinguished triangles.

In the discussion below we will use mirror symmetry to cast some of the
computations in type IIA language, but
we will do so only for simple cases where we can evade
subtle issues of $\Pi$-stability.


\section{The Quantum Mechanics of Special Lagrangians} \label{s:slag}

In this section we will describe the moduli space of BPS D3 branes moving
in the attractor geometry described above.  We will study the moduli space
of probe D3 branes in the geometry produced by a fixed
``background'' D3-brane whose entropy we wish to calculate.  The
probe D3-branes will be taken to be mutually supersymmetric with the
background D3 brane --- they
may be thought of as candidate decay products formed out
of the background D3 branes making up the black hole.
Much of the material in the first part of this section is a straightforward
generalization of \cite{SYZ:mir}.

Consider a stack of $N$ probe D3 branes in the near-horizon $AdS_2
\times S^2 \times Y$ attractor geometry. The probe $D3$ branes are
taken to wrap a special Lagrangian $L\subset Y$, and are point-like
in the $S^2$ and $AdS_2$ spatial directions. Since the $L$
directions are compact, one can integrate over the $L$ directions to
obtain a one-dimensional world volume quantum mechanics. Because of
the $AdS_2$ factor, the theory has an $SU(1,1|2)$ superconformal
symmetry.  Conformal quantum mechanics systems of this type were
described in \cite{GSSY:D0}, which considered D0 branes moving in
IIA attractor geometries. In fact, since our D3 branes are
point-like in the $AdS_2 \times S^2$ directions, these spatial
components of the quantum mechanics are identical to those described
in \cite{GSSY:D0}. We will therefore focus on the Calabi--Yau
component of the moduli space.

Before describing a stack of $N$ D3 branes, first consider a single
A-brane wrapping $L$. First assume we have a {\em smooth embedding\/}
$f: L \to Y$.
The special Lagrangian condition is
\cite{BBS:5b}
\begin{equation}
f^*\omega=0,\qquad
f^*({\rm Im}(e^{-i\xi}\Omega))=0,
\end{equation}
where $\omega $ is the K\"ahler form on $Y$.  In addition, the D3 brane
comes equipped with a world-volume gauge field $A$.\footnote{ This
  world-volume gauge field $A$ should not be confused with the
  connection $\A$ on moduli space that we will discover below.}
Supersymmetry implies that $A$ describes a flat connection on a $U(1)$
bundle\footnote{Assuming $B=0$ on $Y$. A nonzero $B$-field results in
  a ``twisted'' line bundle.} over $L$.  We will now do a local
analysis of the moduli space of these supersymmetric D3 branes,
following \cite{McL:sLag, SYZ:mir} (see also
\cite{Denef:dis,Joyce:2001xt} for a review).

We can imagine deforming a D3 brane in two ways, either by changing $f$ or
by changing $A$.  Infinitesimal deformations of $f$ are in one-to-one
correspondence with harmonic one forms on $L$. To see this, consider a
one-parameter family of embeddings $f_t: {\R} \times L \to Y$ which
preserve the special Lagrangian condition.  Here $t$ is a coordinate on
$\R$. It is straightforward to show that
\begin{equation} f_t^*(\omega) =
\theta \wedge dt, \qquad f_t^*({\rm Im}(e^{-i\xi} \Omega)) = e^{-K/2}
\frac1{2k} *\theta \wedge dt \label{eq:Fcond}
\end{equation}
where $*$ is the Hodge star on $L$ and the constant $k=
\sqrt{8\Vol(Y)}$. Both of these forms are necessarily closed, so
$\theta$ is harmonic.  If we denote by $\theta^a$,
$a=1,\ldots,b_1(L)$, a basis of harmonic one forms on $L$, this
provides us with a family of infinitesimal deformations $dt^a$ of $f$.
It was shown in \cite{McL:sLag} that one can integrate these
infinitesimal deformations to find a good set of local coordinates
$t^a$ on the moduli space of special Lagrangians.

The space of flat connections $A$ is also of dimension $b_1(L)$.  This
is because by a judicious choice of gauge we may always put the
world-volume gauge field in the form $A = \sum_a s^a \theta^a$.  The
constants $s^a$ form a set of coordinates on the moduli space of flat
connections.

It is important to note that the moduli space of A-branes is not
necessarily equal to this moduli space of bundles and special
Lagrangians. Quantum corrections coming from holomorphic disks with
boundary on $L$ can lead to obstructions. Thus, in general, the moduli
space of A-branes can be less than $2b_1(L)$. We refer to
\cite{KKLM:W} for an example of this and \cite{me:TASI-D} for further
discussion.

We will ignore such obstructions here.  We conclude that locally the
moduli space of BPS D3 branes is a product of the form $\M = H^1(L)
\times H^1(L)$, with coordinates $(t^a, s^a)$.  There is a natural
metric on $\M$, of the form $ds^2 = g_{ab} dt^a dt^b + g_{ab} ds^a
ds^b$, where
\begin{equation}
g_{ab} = \frac{1}{2k}\int_L \theta^a \wedge * \theta^b.
\end{equation}
Here, as above, the Hodge star on $L$ is defined using
the metric induced on $L$ by the embedding $f$ of $L$ into $Y$.
Thus $g_{ab}$ is
a function of the embedding $f$, and hence of $t^a$ but not $s^a$.
In fact, the metric can be shown to obey $g_{ab,c} = g_{ac,b}$.
This implies that the natural K\"ahler two form on moduli space,
$\J = g_{ab} dt^a \wedge ds^b$, is closed and defines an integrable complex
structure on $\M$.

We are interested in the dynamics of BPS D3 branes, which are
described by a superconformal quantum mechanics on $\M$.
The kinetic term is found by taking the moduli $(t^a, s^a)$ to depend on time,
and expanding the  Dirac-Born-Infield world-volume
action to quadratic order in $({\dot t}^a, {\dot s}^a)$.  The bosonic
part of the action becomes
\begin{equation}
\begin{split}
S_{DBI} &= \int dt\int_{L} \sqrt{\det(G+B-2\pi \alpha' F)}\\
&= 2k \int dt g_{ab}({\dot t}^a {\dot t}^b + {\dot s}^a {\dot s}^b) + ...
\end{split}
\end{equation}
This is a non-linear sigma model on $\M$.  There are of course also
fermion terms, as well as various terms involving motion in the
$AdS_2$ and $S^2$ directions, which are just as in \cite{GSSY:D0}.
The $2k$ prefactor follows from our choice of normalization of the
metric $g_{ab}$ defined above.  It's importance will become apparent
below, where we discuss Chern-Simons terms which are quantized in
units of $k$.

Now, let us consider what happens for $N$ D3 branes.
The coordinates $(t^a, s^a)$ described above
are promoted to matrices, and the world-volume action will include
terms involving the commutators of these matrices.  For example, there is
now a Chern-Simons type term of the form
\cite{Myers:1999ps, Taylor:1999gq}
\begin{equation}
\int dt \int_L f_t^*(F^{(5)}) [\phi^a,\phi^b]
\label{eq:csnon}
\end{equation}
where $\phi^a$ is an $N \times N$ matrix and $f_t^*(F^{(5)})$ is the
pullback of the Ramond-Ramond field strength (\ref{eq:fsour}) sourced
by the background D3 branes.

Matrix systems of this form admit a large number of possible ground
states, including both commuting and non-commuting configurations of
matrices. According to the Myers effect \cite{Myers:1999ps}, the
various non-commuting configurations should be interpreted as D5 or
D7 branes wrapping cycles in the $S^2 \times X$.  However, a
non-commuting configuration wrapping a cycle in $X$ couples to
Ramond-Ramond fields in the same way as the associated D5 or D7
brane.  It will therefore contribute to the overall charge of the
black hole as measured at infinity. In evaluating the entropy we
should sum only over configurations with the correct asymptotic
charges.  So we should include states where the D3 branes are
allowed to form D5 branes wrapping the $S^2$, but not where they
wrap an internal direction. In this case the matrices $\phi^a$
describing the D3 brane positions form an N dimensional
representation of $SU(2)$.  This representation can be written as a
sum of irreducible representations, each of which corresponds to a
D5 brane wrapping the $S^2$ horizon.  So, the total number of
different ways our N D3 branes may puff up into a collection of D5
branes is equal to the number of partitions of the integer $N$. In
fact, configurations where the D3 branes form a D5 branes in this
way give the dominant contribution to the entropy. This observation
was made in \cite{GSY:bqm}, which studied configurations of D0
branes in IIA that formed a spherical D2 brane wrapping the horizon
$S^2$. \footnote{We should emphasize that the non-commuting
configurations considered here are supersymmetric, as in
\cite{GSSY:D0}, so they are genuine zero energy ground states of the
system.  This is in contrast with the dielectric configurations
originally considered in \cite{Myers:1999ps}, which were
non-supersymmetric and had positive energy.}

For this configuration, where $N$ D3 branes form a D5 wrapping the horizon,
(\ref{eq:csnon}) becomes the usual Chern-Simons interaction term
on the D5 world volume
\begin{equation}
S_{CS} = \int dt \int_{L\times S^2} A \wedge f_t^*(F^{(5)}).
\label{eq:csterm}
\end{equation}
{}From (\ref{eq:Fcond}), the pullback of the associated RR 3-form $F^{(3)}$
onto the brane world-volume is
\begin{equation}
f_t^*(F^{(3)}) =
f_t^*({\rm Im}(C\Omega))
= \frac{|Z|}{\sqrt{8 V}}
\sum_a *\theta^a {\dot t}^a dt.
\end{equation}
In writing the second equality we have used the fact that $\xi=\arg Z$,
since our probe D3-brane is marginally bound to the background D3-brane.
This allows us to compute the pullback using (\ref{eq:Fcond}).
Up to a total derivative, (\ref{eq:csterm}) may be written as
\begin{equation}
\begin{split}
\int dt \int_L F\wedge f_t^*(F^{(3)})
&= \frac{|Z|}{\sqrt{8 V}}
\int dt \sum_{ab} s^a \dot{t}^b \int_L \theta^a\wedge * \theta^b \\
&= |Z| \int dt g_{ab} s^a \dot{t}^b.
\end{split}
\end{equation}
A ``magnetic field'' coming from a $\GU(1)$-bundle with connection
one-form $\A$ contributes a term to the action of the form
\begin{equation}
  S_\A = \int_\gamma \A,
\end{equation}
for a path $\gamma$ in $\M$. Thus we may interpret
the Chern--Simons term as producing a magnetic field with
gauge potential
\begin{equation}
\A = |Z| g_{ab} s^a dt^b.\end{equation}
Now, since $g_{ab,c}dt^b dt^c=0$
we can evaluate the field-strength
\begin{equation}
\F = d\A = |Z| g_{ab} ds^a \wedge dt^b
= |Z| \J .
\label{eq:maga}
\end{equation}


\section{Black Hole Entropy from Mirror Symmetry} \label{s:0B}

We will now use the results of the previous section to compute
the black hole entropy.

We have demonstrated that the dynamics of BPS probe D3 branes is
governed by the world volume quantum mechanics of the form
\begin{equation}
\int dt~{\mathcal{G}}_{ab} \dot{z}^a\dot{z}^b + \A_a \dot{z}^a + {\rm fermions}
\label{eq:scft}
\end{equation}
where $z^a$ and ${\mathcal{G}}_{ab}$ are the coordinates and metric
on moduli space $\M$ of special Lagrangians, and $\A$ is the connection
with curvature $\F \sim \J$ described above.
As this is a superconformal
quantum mechanics, the number of ground states in this system is
encoded in the number of chiral
primaries.  The chiral primary conditions can be written as
\eqn\chp{\bar{D}h=\bar{D}^* h=0, \label{eq:chp}} where
$h$ is a $(p,q)$ form on $\M$
and $D$ is the holomorphic covariant derivative with connection $\A_a$.
Solutions of (\ref{eq:chp}) are in one-to-one correspondence with elements
of $H^{0,q}(X, \Omega^p \otimes \mathcal{L})$.  Here $\mathcal{L}$ is the line
bundle over $X$ with first Chern class $c_1(\mathcal{L})=[\F]$.
So the black hole entropy counting
can be reduced to a cohomology problem.

We should emphasize that
in order to render the supergravity approximation implicit in
the previous discussion valid, we must consider black holes with
large charge. This makes the
general computation of the cohomology an exceedingly formidable
task. Even relatively simple D-branes such as certain ones on the
quintic threefold are hard to study in terms of $\Pi$-stability
\cite{AD:Dstab}. Added to this is the complication that we
need to examine an enormous number of possible decay paths as
discussed in the previous section.

However, as we describe in the next section, in some cases one can
compute the dimension of $H^q(X, \Omega^p \otimes \mathcal{L})$
using mirror symmetry.
As an illustration, we will show how this works for the simple
case $Y=T^6$.

\subsection{Mirror Computation}

There is, of course, one 3-brane for which the moduli space is easy to
compute. If $X$ is mirror to $Y$, then we know from homological mirror
symmetry (or, less rigorously but more transparently, by using
\cite{SYZ:mir}) that a 0-brane on $X$ is mirror to a 3-torus on $Y$.
The moduli space of a D3-brane wrapping such a $T^3$ on $Y$ is just $X$.
Thus one can compute
$H^q(X, \Omega^p \otimes \mathcal{L})$ ---
and hence the black hole entropy --- using mirror symmetry.
We will consider a black hole whose charge is dominated by such
3-branes.

Consider a B-brane on a \CY\ threefold $X$ whose charge is large and
made up almost entirely of 0-branes. Let us also assume that a
0-brane is a possible decay product among the multitude of possible
marginal decays. This means that the inner product under
(\ref{eq:Binr}) of a 0-brane with our black hole brane $L$ must be
zero. If $E$ is a 0-brane, then $\ch(E)$ is a pure 6-form. Thus,
using (\ref{eq:Binr}) we require that $\ch(L)$ has no 0-form part.
The 0-form part of a Chern character measures the rank of a vector
bundle. Thus, our big D-brane must correspond to a vector bundle of
rank 0, i.e., it is supported over a proper holomorphic subspace of
$X$. To put it another way, {\em it must have no 6-brane
  charge.}

The probe
quantum mechanics of the special Lagrangian 3-brane described in the
previous section can now be recast as the quantum mechanics of D0 branes
on $X$.  This D0 brane theory
was studied in \cite{GSSY:D0}, and used to compute the black hole entropy
in \cite{GSY:bqm}. The rest of this subsection is essentially a review
of this work, which we include here to make the discussion self contained.


We should emphasize that we cannot consider black holes made entirely
of D0 branes, however, as in the leading supergravity approximation these
black holes have zero area.  So we need to include some 4-brane charge.
This means that there will be a moduli space of states
involving 4-branes.  However, we should note that the
number of 0-branes must be much greater than the number four branes wrapping
any given cycle in $X$.  To see this, note that
the K\"ahler form on $X$ (in ten dimensional units) is determined
by the IIA
attractor equation to be
\begin{equation}
J = \sqrt{\frac{q_0}{D}} p^A \omega_A, ~~~~~
D = D_{ABC} p^A p^B p^C.
\label{eq:twocycle}
\end{equation}
Here we have chosen a basis $\omega_A$ of $H^2(X,\bf{Z})$ and denoted by
$p^A$ the number of $4$-branes wrapping the 4-cycle Poincar\'e dual
to $\omega_A$.  Here
$D_{ABC} = \ff16 \int \omega_A \wedge \omega_B \wedge \omega_C$
is a triple intersection number and $q_0$ is the number of 0-branes.
As discussed in section 2.1,
the supergravity approximation is valid only when the size of any two
cycle in $X$ is much smaller than the black hole horizon area (although it
must still be large in string units).  From (\ref{eq:twocycle}), this
implies
that we must have $q_0 \gg p^A$.
As long as the 0-brane charge
dominates, the contributions from the moduli space of 4-branes should
be negligible in determining the entropy.%
\footnote{However, there has been recent progress in understanding
these 4-brane contributions in some cases \cite{Gaiotto:2005rp}.} As
discussed above, while it is in principle possible to add D2 brane
charge to such a black hole, one can not add D6 branes and maintain
supersymmetry.

Now, the moduli space of a 0-brane moving on $X$ is just $X$ itself.
This means that the metric ${\mathcal G}_{ab}$ described in section
\ref{s:slag} is just the metric on $X$.  In section \ref{s:slag} we
demonstrated that there is magnetic field on moduli space whose
field strength is equal to the central charge $|Z|$ times the
K\"ahler form on $X$ given above. In the IIA language, it is easy to
see where this magnetic field comes from.  The IIA supergravity
solution for the black hole includes a Ramond-Ramond four form
fieldstrength $ \omega_{S^2} \wedge p^A \omega_A$, which is sourced
by the background D4 branes. For configurations of fuzzy D0 branes
wrapping the $S^2$ horizon, this four form fieldstrength couples to
the D0 brane worldvolume fields via the Myers effect, in a manner
similar to that describe in section 3.  The result is a magnetic
fieldstrength $ \F = p^A\omega_A$ on moduli space
\cite{GSY:bqm}.\footnote{This expression for $\F$ also follows from
equations (23) and (26).  However, one must be careful because the
usual expression for central charge $|Z|=\left(q_0 D\right)^{1/4}$
is written in four dimensional Planck units rather than in ten
dimensional units.  To rewrite it in ten dimensional units we must
use the conversion factor
\begin{equation}
\frac{l_4}{l_{10}}=l_{10}^3 \Vol(X)^{-1/2} \label{eq:scale}.
\end{equation}
where $\Vol(X) = \int_X J^3 = q_0^{3/2} D^{-1/2}$.}

To compute the cohomology, note that the bundle $\mathcal{L}$ is very ample (as
its first Chern class is the K\"ahler form).
Thus $H^{p,q}(X,\mathcal{L})=0$ if $p>0$.
We can then use an index theorem to compute $H^{0,q}(X, {\mathcal L})$
just as in \cite{GSY:bqm}.  To leading order in the charges, the answer is
\begin{equation}
\dim H^{0,q}(X,\mathcal{L}) = \begin{cases}
D&q=0,3,\\
3D&q=1,2.
\end{cases}
\end{equation}

We now need to count the total number of supersymmetric
ground states of the theory, using the chiral primary degeneracies
$h^q = \dim(H^{0,q} (X, {\mathcal L}))$
computed above.  As usual, it is convenient to package the answer as
a partition function
\begin{equation}
Z(q) = \sum_N p(N)q^N .
\end{equation}
Here $p(N)$ is the number of chiral primary states that may be
formed out of $N$ D0 branes.  As described in section 3, according
to the Myers effect these $N$ D0 branes may form a collection of D2
branes wrapping the horizon $S^2$.  There is one such collection of
D2 branes for each partition of the integer $N$.  In addition, each
D2 brane so formed may occupy any one of the chiral primary states
on the Calabi--Yau counted above. The only restriction is that chiral
primaries with even $q$ obey bosonic statistics, while states with
odd $p$ obey fermionic statistics. Putting this together, it turns
out that $Z(q)$ is precisely the partition function of a conformal
field theory with $h^0 + h^2$ free bosons and $h^1 + h^3$ free
fermions:
\begin{equation}
Z(q) = \prod_n\frac{(1+q^n)^{h_1 + h_3}}{(1-q^n)^{h_0+h_2}} .
\end{equation}
Using the large $N$ expansion for this formula gives the entropy
\begin{equation}
\begin{split}
  S = \log p(N) &\sim
    \pi\sqrt{\ff13 N \left(h^0 + h^2 + \ff12(h^1 + h^3)\right)}.
\sim
2\pi \sqrt{N D}
\end{split}
\end{equation}

As a word of warning,
one should probably view the computation in this section and the previous
section as somewhat
heuristic, as it is prone to the same objections as the general SYZ
argument. We have certainly not taken into account the fact that
special Lagrangians typically degenerate for certain points in
the moduli space. However, we will see that we can
reconstruct this result more rigorously for
K3$\times T^2$ in section \ref{s:K3}.

\subsection{An $\N = 8$ Example} \label{ss:T6}

In this section we will describe the computation for the case
$Y=T^6$, which provides a useful explicit
illustration of these methods.
Here the moduli space of special
Lagrangians takes a particularly simple form and the calculations
can be done explicitly without using the mirror symmetry conjecture of
\cite{SYZ:mir}.

It is straightforward to write down and solve the attractor equations on $T^6$
(see e.g. \cite{Moore:arith}).
We will take coordinates on the $T^6$ to be
$x^i\sim x^i+1, y^i \sim y^i + 1$, $i=1,2,3$.
A choice of holomorphic one forms $dz^i = dx^i + \tau^{ij} d y^j$  fixes the
period matrix $\tau$.
We will suppress $i$ indices when possible.
The metric is
\eqn\asd{
dz \cdot d{\bar z} = dx \cdot dx + 2 dx \cdot Re \tau \cdot dy +
dy \cdot \tau^\dagger \tau \cdot dy
.}
We will take the following symplectic basis for
$H^3(T^6,{\Z})$,
\begin{equation}\begin{split}
\alpha_0 &= dx^1 dx^2 dx^2,~~~~~\alpha_{ij} =\half \epsilon_{ilm} dx^l dx^m dy^j \\
\beta^0 &= - dy^1 dy^2 dy^2,~~~~~\beta^{ij} =\half \epsilon_{jlm} dx^i dy^l dy^m
\end{split}\end{equation}
so that $\int \alpha_I \wedge \beta^J = \delta_I^J$, where $I=(0,ij)$.
The charge of the black hole is parameterized by an element of
$H^3(T^6)$, which can be decomposed in this basis as
\eqn\asd{
F^{(3)} = p^0 \alpha_0 + P^{ij} \alpha_{ij} + q_0 \beta^0 + Q_{ij} \beta^{ij}
.}
We will focus on the case where $p^0$ and $Q_{ij}$ vanish and $P$ is
symmetric. In this case the attractor equations fix the holomorphic three form
to be (up to an overall constant)
\begin{equation}
\Omega^{3,0} = dz^1 \wedge dz^2 \wedge dz^3
= \alpha_0 + \alpha_{ij} \tau^{ij} + \beta^{ij}
(\operatorname{Cof} \tau)_{ij} - \beta^0 (\det \tau)
\end{equation}
where $(\operatorname{Cof} \tau)_{ij}$ is the matrix of cofactors of
$\tau^{ij}$ and
\eqn\asd{
\tau = \ff{i}2 \sqrt{\frac{q_0}{\det P}} P
.}
The entropy of the black hole is
\eqn\toruss{
S = 2 \pi\sqrt{ |q_0  \det P| }. \label{eq:toruss}}

Now, consider a probe three brane wrapping $\alpha_0$
(i.e. the $x^i$ directions).  The induced world-volume metric is flat.
The moduli space is parameterized by a position $y^i(t)$ and
world-volume gauge field $A = a_i(t) dx^i$, which depend on time.
The kinetic terms for $y^i$ and $a_i$ come from the DBI action
\eqn\asd{
S_{DBI} = \int dt
\left({\dot y} \cdot \tau^\dagger \tau \cdot {\dot y} + {\dot a} \cdot {\dot a}
\right) + ...}
The world volume Chern-Simons term depends only on $P^{ij}$, and is
\eqn\asd{
S_{CS} = \int A \wedge f^* F_3 = \int dt
\left(a \cdot P \cdot {\dot y} \right)
.}
The action $S_{DBI} + S_{CS}$ describes a particle in a magnetic field,
with moduli space metric and magnetic field strength
\eqn\asd{
ds^2 = dy \cdot \tau^\dagger \tau \cdot dy + da \cdot da,
\ \ \ \ \ \ \
\F = dy \cdot P \cdot da
.}
Note that both the $a$ and $y$ coordinates are identified with
periodicity one.
As an aside, we should note that the two form
$\F$ defines a complex structure on the
moduli space.  The associated holomorphic coordinates are
\eqn\asd{
w= a + \tau \cdot y  ,
}
in terms of which the metric and field strength are
\eqn\asd{
ds^2 = dw \cdot d{\bar w},
\ \ \ \ \ \ \
\F = - i \sqrt{\frac{\det P}{q_0}}  dw \cdot \wedge d{\bar w}
.}

We can now count the number of chiral primaries, by computing
$H^{0,p}(T^6,\Omega^q\otimes \mathcal{L}$) where $\mathcal{L}$ is a
line bundle with curvature $c_1(\mathcal{L}) = \F$.
By a Kodaira vanishing theorem, these vanish unless $p=0$
(at large charge) so we may use
an index theorem for the twisted Dolbeault complex:
\begin{equation}
h^{0,q} = \textrm{Ind}_{\bar \partial}(T^6, \Omega^q\otimes \mathcal{L}) 
= \binom3q\int \F\wedge \F \wedge \F
= \binom3q \det P .
\end{equation}
We can then count the number of chiral primaries using the combinatorics of
the previous section, reproducing the black hole entropy (\ref{eq:toruss}).

\section{An $\N=4$ Example} \label{s:K3}

In this section we will consider compactifications with $\N=4$
supersymmetry, where $Y$ is of the form $T^2\times S$ for some K3
surface $S$.  Although in this case the moduli space of special
Lagrangians is more complicated than the $\N=8$ example
described above, we may still perform the computation without
relying on the analysis of section
\ref{s:slag}. One may view this section as evidence that
(\ref{eq:maga}) is correct even when the special Lagrangian fibration
in section \ref{s:slag} has degenerate fibers. Some of what we discuss
below is related to the ``Donaldson--Mukai map'' as described in
\cite{Dijkgraaf:1998gf}.

The attractor equations for $T^2 \times K3$ were studied in
\cite{Moore:arith}; we will simply quote the results here.
We will denote the coordinates on $T^2$ as $x$ and $y$,
with
$x\sim x+1$ and $y\sim y+1$. The charge of the black hole is
\begin{equation}
W = p\,dx + q\,dy
\end{equation}
where $p$ and $q$ live in $H^2(S,\Z)$.  We will write the holomorphic
3-form on $Y$ as
\begin{equation}
\begin{split}
\Omega^{3,0} &= dz\wedge\Omega\\
&= (dx + \tau\,dy)\wedge\Omega,
\end{split}
\end{equation}
where $\tau$ is the complex modulus of $T^2$ and
$\Omega$ is a holomorphic 2-form on $S$.
The attractor equation
(\ref{eq:att}) then forces (for a suitable choice of normalization)
\begin{equation}
\Omega = q-\bar\tau p.
\end{equation}
Furthermore, $\tau$ is determined by $p$ and $q$ to be
\begin{equation}
  \tau = \frac{p\cdot q+i\sqrt{p^2q^2-(p\cdot q)^2}}{p^2},  \label{eq:tau}
\end{equation}
where the dot product is given by the usual intersection form on $S$:
\begin{equation}
  p\cdot q = \int_S p\wedge q.
\end{equation}

We will now consider a probe 3-brane on $Y$, and interpret
the Chern--Simons contribution to the world volume action as
an effective magnetic field on moduli space, which in this case
will be the mirror, $X$, of $Y$.

We will take the probe to wrap the
special Lagrangian cycle in $T^2\times S$
that consists of the circle Poincar\'e dual to $dy$ times a
special Lagrangian 2-cycle $L\subset S$. As above, this probe D-brane must
have its phase aligned with that of the background black hole, so
from (\ref{eq:align})
\begin{equation}
\begin{split}
  \int_{Y^2\times S} (p\,dx + y\,dy)\wedge l\,dy &= l\cdot p\\
  &= 0,
\end{split} \label{eq:ldotp}
\end{equation}
where $l$ is the 2-form Poincar\'e dual to $L$.  Since $L$ is
Lagrangian, $l\cdot J=0$ where $J$ is the cohomology class of the
K\"ahler form on $S$.

To write down the Chern--Simons term, we need to consider a one
parameter family of D-brane probes.  For now let us focus on just the
K3 factor and consider a one-parameter family of maps $L\to S$.  We
will take this one-parameter family to form a loop --- i.e., the final
D-brane is the same as the initial D-brane.  Denote by $T$ the
3-dimensional subspace swept out in $S$ by this family; note that $T$
has no boundary, since the one parameter family is a loop.  Each
special Lagrangian D-brane comes equipped with a line bundle and a
connection $A$, which we will extend to a connection over $T$.  The
Chern--Simons term is just\footnote{We are assuming this family
  produces an {\em embedding\/} of $T$ into $S$. This need not be the
  case in general but is true for the case we consider here.}
\begin{equation}
S_{CS} = \int_T A\wedge p.
\end{equation}
Now, note that the third homology of
$S$ is zero, so there must be a 4-dimensional subspace $U\subset S$
such that $T=\partial U$. Stokes' theorem yields
\begin{equation}
S_{CS} = \int_U F\wedge p.\label{eq:CSS}
\end{equation}
Let $\M_S$ be the moduli space of probe D2 branes on $S$, i.e., the moduli
space of special Lagrangians with bundle data.
Recall that $iS_{CS}$ takes values in $\R/2\pi\Z$.
It therefore defines a map
\begin{equation}
  S_{CS}:\Omega(\M_S)\to S^1,  \label{eq:CSm1}
\end{equation}
where $\Omega(\M_S)$ is the loop-space of $\M_S$. Given any space $M$,
$\pi_2(M)$ is defined as $\pi_1(\Omega(M))$. So, applying $\pi_1$ to
(\ref{eq:CSm1}) yields a map
\begin{equation}
  \eta_{CS}:\pi_2(\M_S)\to \Z.
\end{equation}

The moduli space $\M_S$ is expected to break up into many disconnected
components, corresponding to different types of probe D-branes.  To
proceed, we must therefore choose the type of probe brane under
consideration.  As above we will assert that $L$ is a $T^2$-fiber of
$S$ so that it is mirror to a 0-brane.  According to the SYZ
conjecture \cite{SYZ:mir}, which is easily proven for K3 (as we
discuss next), this component of $\M_S$ is itself a K3 surface. We
denote this component $\hat S$. In a sense, $\hat S$ is ``the'' mirror
of $S$.

The rigorous way to prove that $\hat S$ is a K3 surface is as follows.
The K3 surface $S$ is hyperk\"ahler, and so admits a whole $S^2$ of
complex structures compatible with a given metric. By choosing a
different complex structure, we may turn a special
Lagrangian into a holomorphic curve. To be
precise, we may turn the special Lagrangian 2-torus with a flat line bundle
into an elliptic curve in $S$ with a flat line bundle. This
is a sheaf supported on the elliptic curve. One may then compute
the moduli space $\hat S$ of such sheaves. This was done by Mukai
\cite{Muk:K3}, who showed that $\hat S$ is a K3 surface.
In fact, $\hat S$ is isomorphic to $S$ as a complex variety.

It is worth emphasizing that attractive K3 surfaces have many special
properties. Because of their large Picard number, they contain many
algebraic curves and thus homologically distinct elliptic curves. Any
elliptic curve leads to an elliptic fibration so a typical attractive
K3 surface can probably be elliptically fibered in many inequivalent
ways. It is always possible to choose an elliptic fibration of an
attractive K3 surface such that it admits a {\em section\/}. The
explicit form of such a fibration was given in \cite{SI:singK3}. We
will assume that we have chosen the fibration so that this is the case.

Consider this section $\sigma$ of the elliptic fibration $\hat S$. This
corresponds to an element of $\pi_2(\hat S)$. The precise family of
elliptic curves (or special Lagrangians, depending on the chosen
complex structure) in $S$ corresponding to this section may be determined by
following Mukai's construction. Indeed, a Fourier--Mukai transform may
be applied to this section to yield the family of elliptic curves on
$S$. For an account of this transform in this context we refer to
\cite{AD:tang}.

In our case, this family of elliptic curves will sweep out a line
bundle $E$ over the whole of $S$. This line bundle is, of course, the
bundle whose curvature $F$ appears in (\ref{eq:CSS}). It follows that
\begin{equation}
\eta_{CS}(\sigma) = \frac1{2\pi i}\int_S F\wedge p.
\label{eq:etacs}
\end{equation}
We can now use the Fourier-Mukai transform to express this
Chern--Simons term as a gauge connection on the moduli space $\hat S$.
To do this, we will first write (\ref{eq:etacs}) in a more convenient
form.  To start, recall from section \ref{s:decay} that there is a
natural inner product on $H^*(S,\Z)$,
\begin{equation}
  \langle \alpha,\beta\rangle = \int_S \alpha^\vee\wedge\beta
\end{equation}
for any $\alpha$ and $\beta$ in $H^*(S,\Z)$.  Here $\alpha^\vee$ is
just $\alpha$ with the sign of the 2-form component reversed.
Furthermore, to any bundle (or sheaf) we may associate its D-brane
charge, or Mukai vector, defined as
\begin{equation}
  v(E) = \ch(E)\wedge\sqrt{\td(T_S)}\in H^*(S,\Z).
\end{equation}
We may therefore write
\begin{equation}
  \eta_{CS}(\sigma) = -\langle v(E),p\rangle.
\end{equation}

Mukai's mirror symmetry construction can now be applied using the
Fourier--Mukai transform. This has some known action on $H^*$,
which we denote $\mu$:
\begin{equation}
\mu:H^*(S,\R) \to H^*(\hat S,\R).
\end{equation}
This action has the nice feature that it preserves the inner product
$\langle\alpha,\beta\rangle$. Thus
\begin{equation}
  \eta_{CS} = -\langle v(\sigma),\mu(p)\rangle.  \label{eq:CSm}
\end{equation}
We now just have to evaluate $\mu(p)$.
This will require a few more facts about K3 surfaces.

As is standard in the construction of string theories on K3
(see, for example, \cite{me:lK3}) the moduli of $S$ are determined by a
space-like 4-plane $\Pi\subset\R^{4,20}$, where
$\R^{4,20}=H^*(S,\R)$. To put a geometric interpretation on this
4-plane one proceeds as follows:
\begin{enumerate}
\item First one chooses a vector $w$ in the
lattice $H^*(S,\Z)$ which generates $H^4(S,\Z)$, and a vector $w^\vee$
which generates $H^0(S,\Z)$. We then identify $H^2(S,\Z)$ as the
orthogonal complement of the span of $w$ and $w^\vee$. Note that
\begin{equation}
  \langle w,w\rangle=0,\quad \langle w,w^\vee\rangle=1,
       \quad \langle w^\vee,w^\vee\rangle=0.
\end{equation}
\item Define $\Sigma'$ = $\Pi\cap w^\perp$.
\item Define the vector $x$ such that $\Pi$ is the span of $\Sigma'$
  and $x$, $x$ is orthogonal to $\Sigma'$ and $\langle
  x,w\rangle=1$.
\item Project $\Sigma'$ into $H^2(S,\R)$ to obtain $\Sigma$.
\item Decompose
\begin{equation}
   x = \alpha w + w^\vee + B,
\end{equation}
where $B\in H^2(S,R)$.
\end{enumerate}

$\Sigma$, a space-like 3-plane in $H^2(S,\R)$, is then spanned by
$\Re(\Omega)$, $\Im(\Omega)$ and $J$.  In fact, the 2-plane spanned by
$\Re(\Omega)$ and $\Im(\Omega)$ is spanned by $p$ and $q$, since we
have assumed that $S$ is an attractive K3 surface.  So $\Sigma$ is
spanned by $p$, $q$ and $J$.

The $B$-field is then given by $B$; and $\langle J,J\rangle$, the
volume of the K3 surface, is given by $\langle x,x\rangle=2\alpha +
\langle B,B\rangle$.

Now, $l$ --- the Poincar\'e dual of our special Lagrangian fiber --- is
perpendicular to both $p$ (from (\ref{eq:ldotp})) and $J$ (since it is
Lagrangian). We can now apply mirror symmetry, which consists of the
hyperk\"ahler rotation of the complex structure required to make
the fiber holomorphic, followed by the Fourier--Mukai transform
$\mu$. Our special Lagrangian fiber turns into a 0-brane, so
$\mu(l)$ must be a 4-form, and thus $w$. Now, since $l$ is
perpendicular to $p$, $\mu(p)$ is perpendicular to $w$.

To fix $\mu(p)$ completely we must deal with one subtlety.  Mirror
symmetry is generally thought of as exchanging deformations of complex
structure with deformations of $B+iJ$. This is somewhat ambiguous for
K3 surfaces, because the hyperk\"ahler structure provides many
possible maps that can be interpreted as mirror symmetry.  This
ambiguity may be fixed as follows.  We will take the space-like
4-plane $\Pi$ to be spanned by two 2-planes, $\Omega$ and $\mho$,
where $\Omega$ is spanned by $p$ and $q$ and $\mho$ is spanned by $J'$
and $x$.\footnote{We are implicitly assuming that $B$ lies in the span
  of {\em algebraic\/} 2-cycles. We are free to make this assumption
  for attractive K3 surfaces.}  $J'$ is projected into $H^2(S,\R)$ to
obtain $J$.  $\Omega$ and $\mho$ should be thought of as encoding
complex structure and $B+iJ$ deformations, respectively.  We now
require that mirror symmetry interchanges $\Omega$ and $\mho$.

We will use hats to refer to quantities for $\hat S$. Thus
\begin{equation}
\mu(\Omega)=\hat\mho,\qquad \mu(\mho) = \hat\Omega.
\end{equation}
$\mu(p)$ must lie in the plane spanned by $\hat J'$ and $\hat x$.
Since $l$ is perpendicular to $p$ and not to $q$, $\hat w=\mu(l)$ is
perpendicular to $\mu(p)$ and not to $\mu(q)$.  But $\hat w$ is
perpendicular to $\hat J'$ and not $\hat x$, so
\begin{equation}
  \mu(p) = \hat \kappa J',
\end{equation}
for some real number $\kappa$. We may assume that we have chosen our
mirror symmetry transform such that $\kappa>0$.

$\kappa$ may be computed from above by knowing the volume of $\hat
S$. We compute
\begin{equation}
  \hat x = \frac1{(q.l)}\left(q-\frac{(p.q)}{p^2}p\right).
\end{equation}
It follows from above that
\begin{equation}
\begin{split}
 \hat J^2 &= \hat x^2\\
    &= \frac{p^2 q^2 - (p.q)^2}{p^2(q.l)^2}
\end{split}
\end{equation}
Thus, using (\ref{eq:tau}), we have
\begin{equation}
  \kappa = \frac{(q.l)}{\Im(\tau)}.
\end{equation}

We conclude that the Chern-Simons contribution  (\ref{eq:etacs}) is simply
related to $\hat J'$:
\begin{equation}
\begin{split}
  \eta_{CS} &= -\langle v(\sigma),\mu(p)\rangle\\
    &= \int_S \check\sigma\wedge \kappa \hat J'\\
    &= \kappa \int_\sigma \hat J',
\end{split}
\end{equation}
where $\check\sigma$ is Poincar\'e dual to $\sigma$. Note that $\hat
J'=\hat J+r\hat w$ for some real number $r$. The $r\hat w$ component
of $\hat J'$ corresponds to a 4-form and therefore has no contribution
in the above. We can therefore assert that
\begin{equation}
  \eta_{CS} = \kappa \int_\sigma \hat J.
  \label{eq:etaCS}
\end{equation}

In fact, this is precisely the contribution of a magnetic field
on moduli space, with curvature equal to $\kappa\hat J$.  To see this,
consider a connection $\mathcal{A}$ on $\hat S$, and let $\gamma$ be a
loop in $\hat S$. Since $\pi_1(\hat S)=0$, there is a disk $D$ such
that $\partial D=\gamma$. The Wilson line associated to this loop
contributes to the path integral as
\begin{equation}
S_W = \int_\gamma \mathcal{A} = \int_D \mathcal{F},
\end{equation}
where $\mathcal{F}$ is the curvature of $\mathcal{A}$. This time $iS_W$ is
valued in $\R/2\pi\Z$ and so we may repeat the argument given
above to get a map
\begin{equation}
\eta_W:\pi_2(\hat S) \to \Z,
\end{equation}
where, for our section $\sigma$ we have
\begin{equation}
\eta_W(\sigma) = \frac1{2\pi i}\int_\sigma\mathcal{F}.
\end{equation}
Comparing this to (\ref{eq:etaCS}) we see that  the
  effect of the Chern--Simons term on $S$ is mirror to the Wilson
  line contribution of a connection on a line bundle on $\hat S$ whose
  curvature is given by $2\pi i\kappa\hat J$.

We may also analyze the contribution to the magnetic field from the
  $T^2$ part of $Y$. This computation can be done in a way similar to
  section \ref{ss:T6}. The result is that the cohomology class of the
  curvature obeys
\begin{equation}
  \left[\ff{1}{2\pi i}\F\right] = \mu(p) + (q.l)e,
\end{equation}
where $e$ generates $H^2(T^2,\Z)$ and the K\"ahler form is given by
(at least in cohomology)
\begin{equation}
  \hat J = \frac{\Im(\tau)}{(q.l)}\left[\ff{1}{2\pi i}\F\right].
\end{equation}
This is the analogue of equation (\ref{eq:maga}) subject to the same
scale change as (\ref{eq:scale}).

The area of the event horizon can be computed \cite{Moore:arith} to be
$4\pi\Delta$, where
\begin{equation}
  \Delta = \sqrt{p^2q^2-(p.q)^2}.
\end{equation}

So far everything we have said in this section is exact. In order to
compute the entropy we need to start making some approximations.
Let us decompose $p$ and $q$ as follows:
\begin{equation}
\begin{split}
  p &= sl + \tilde p\\
  q &= Nl + ml^\vee + \tilde q,
\end{split}
\end{equation}
where $l^\vee$ is Poincar\'e dual to the sum of the section and fiber
of the elliptic fibration (i.e., $\mu(l^\vee)=w^\vee$), $\tilde q$
is perpendicular to $l$ and $l^\vee$ and similarly for $\tilde p$.

Following the mirror symmetry construction above one finds that on
$X=\hat S\times T^2$ we have the following interpretation of these
charges:
\begin{itemize}
\item $N$ counts the 0-brane charge.
\item $\tilde q$
counts 2-branes wrapping 2-cycles in $\hat S$.
\item $s$ counts 2-branes wrapping the $T^2$ factor.
\item $\tilde p$ counts 4-branes wrapping the $T^2$ factor and 2-cycles in $\hat
S$.
\item $m$ counts 4-branes wrapping $\hat S$.
\item As promised in section \ref{s:0B} there can be no 6-brane
charge.
\end{itemize}

Since we are assuming 0-brane charge dominates, we have $q^2\simeq
2Nm$ and $p^2q^2\gg (p.q)^2$. This gives
\begin{equation}
  \textrm{Area} = 4\pi\sqrt{2Nmp^2}.  \label{eq:K3area}
\end{equation}

Now we can compute the entropy using the method of section \ref{s:0B}.
As in \cite{GSY:bqm} we may use an index theorem to compute the Hodge
numbers $h^{p,0}$ in terms of $\F$. We obtain
\begin{equation}
  S = \pi\sqrt{2Nm(p^2-\ff23)}.   \label{eq:K3ent}
\end{equation}
Thus we have agreement between the area and entropy so long as $p^2\gg 1$.
This condition is explained by the discussion at the end of section
\ref{ss:attr}. It is easy to compute the area of the $T^2$ factor:
\begin{equation}
\begin{split}
  \Area(T^2) &= \int_{T^2} \hat J\\
  &= \frac{\Delta}{p^2}.
\end{split}
\end{equation}
Thus the ratio of the area of event horizon to the area of the $T^2$
factor, which must be large, is $4\pi p^2$. If we view this as a type
IIB compactification, then the complex structure of $T^2$ is
constrained to not be too large as expected in section \ref{ss:attr}.

Note that the 2-brane charge, $\tilde q$ and $s$, plays no r\^ole in
either the entropy or the area so long as the D-brane is dominated by
0-brane charge. If $Nm$ is not much greater than $\tilde q^2$, for
example, then we will have a non-negligible 2-brane contribution to
the area and (\ref{eq:K3area}) will not longer be valid. However, in
this case we will have other important decay modes involving 2-branes
that we have not accounted for and so (\ref{eq:K3ent}) will be
modified too.

\section{Conclusion}

We have described a procedure for computing the entropy of
Calabi--Yau black holes in type IIB string theory, at least for
large charges. This procedure is simple to state, but in practice is
technically challenging. It relies on an ability to analyze the
stability of the given D-brane and enumerate the resulting
constituent D-branes.  In addition, one requires an understanding of
the moduli space of special Lagrangians, which we can claim only for
a small subsector. Within that context, we have provided evidence
that it works for some simple examples. It should also be noted that
this procedure highlights the special nature of Calabi--Yau spaces
at attractor points. The analysis relies crucially on the
near-horizon superconformal quantum mechanics developed by
\cite{GSSY:D0, GSY:bqm}.

Clearly it would be nice to go beyond the examples described in this paper,
and compute the entropy of D-branes dominated
by some charge other than 0-branes (or their mirrors).
That is, we would like to consider probes that are not mirror to 0-branes.
Section \ref{s:slag} indicates that this is possible.
We argued that the moduli space of probe branes is valued in a line
bundle $\mathcal{L}$ over the moduli space of special Lagrangians,
with $c_1(\mathcal{L})$ given by
the K\"ahler form associated to the natural metric on
this moduli space. Although the methods in section \ref{s:slag} cannot
be considered rigorous, it is certainly tempting to conjecture that
this beautifully simple result is true in general. Counting the probe
quantum states is then given by cohomology with values in this bundle.

An obvious place to look for other examples is $T^2\times\mathrm{K3}$.
The probe 3-brane will be of the form $S^1\times C$ for some 2-cycle
$C$ in K3.  In section 5 we described the case where $C$ is an
elliptic curve. One might think the simplest case to consider is
$C\cong S^2$. The moduli space of such a curve is a point, so the
analysis starts to look rather trivial.  In fact, for black holes
whose charge is dominated by such a D-branes it is easy to prove that
$p^2q^2-(p.q)^2 < 0$.  This means that the attractor mechanism breaks
down and the solution is not a spherically symmetric black hole.

The next case to consider would be a curve $C$ of genus $g>1$. The
moduli space of such curves is understood to an extent following
Mukai's work \cite{Muk:symp}. In particular it is known that the
moduli space is hyperk\"ahler and of complex dimension $2g$.  This was
studied in \cite{Dijkgraaf:1998gf}.  It would be interesting to see if
some examples in this case could be computed.

A few caveats remain. We have not explained the insight of
\cite{GSY:bqm} that the dominant contribution to the entropy comes
from states which wrap the horizon a single time.
We also have not provided a guess as
to the states which give the subleading terms to the entropy (as an
expansion in inverse charge). The answers to these questions
will surely lead to new insights into the nature of black holes
in string theory.


\section*{Acknowledgments}

We wish to thank A.~Adams, C.~Beasley, D.~Gaiotto, J.~Hsu,
R.~Kallosh, S.~Kachru, A.~Kashani-Poor, A.~Strominger, A.~Tomasiello
and X.~Yin for discussions. A.~M. and A.~S. wish to thank the Center
for Mathematical Sciences, Hangzhou, Zhejiang, China, for
hospitality during the initial stages of this work. P.S.A.~is
supported by NSF grants DMS--0301476 and DMS--0606578. The work of
A.~M. is supported by the Department of Energy, under contracts
DE--AC02--76SF00515 and DE--FG02--90ER40542.


\end{document}